\documentclass[12pt]{article}
\date{}
\usepackage{hyperref}
\usepackage{xcolor}

\usepackage{placeins}
\usepackage[latin1]{inputenc}
\usepackage{amsmath}
\usepackage{amssymb,amsthm}
\usepackage[english]{babel}
\usepackage[textwidth=6.25in,textheight=8.39in]{geometry}
\usepackage{setspace}
\usepackage{tabularx,ragged2e,booktabs,caption}
\usepackage[toc,page]{appendix}
\usepackage[affil-it]{authblk}
\usepackage{graphicx}
\graphicspath{{/imgs/}} 
\newcolumntype{C}[1]{>{\Centering}m{#1}}

\begin{document}

\title{Digital Archives as Big Data}
\author{Luis Martinez-Uribe}
\affil{Departamento de Sociolog\'{\i}a, Universidad de Salamanca\\
DataLab, Fundaci\'on Juan March, Calle Castello 77, Madrid\\
email: lmartinez@march.es}

\maketitle \abstract{ \vspace*{.5cm}

\begin{center}\begin{minipage}{12cm}

Digital archives contribute to Big Data. Combining social network analysis, coincidence analysis, data reduction, and visual analytics leads to better characterize topics over time, publishers´ main themes and best authors of all times, according to the British newspaper \textit{The Guardian} and from the 3 million records of the British National bibliography.

\vspace{5mm} 

\emph{Keywords:} Big Data, coincidence analysis, social network analysis, open data.

\end{minipage}
\end{center}
\newpage

\section{Introduction}
Latour´s (2007: p.2) quote: \textit{``It is as if the inner workings of private worlds have been pried open because their inputs and outputs have become thoroughly traceable''},  summmarizes Big Data.

Big Data comprise not only government databases, social media, transactions such as credit cards and online clicks, and global positioning systems or accelerometers but also stem digitized documents from libraries and archives (Martinez-Uribe and Fernandez, 2015). 

In bibliometrics, clustering, time series, and network analysis have already been used to rate citations and identify production and co-authorship (Korenjak-Cerne et al., 2006; Battisti and Salini, 2012; Ferrara and Salini, 2012). Library resources are of high quality (Topçu et al., 2014), be it in literature (Moretti 2005), history (Cohen, 2006), musicology (Tuppen et al., 2016), or sociology (Escobar, 2009; Escobar and Isla, 2015).

We show what kind of insight bring graphics and visualization (Healy and Moody, 2014; Cook et al., 2016)by combining coincidence analysis, data reduction, social network, and visual analytics. We treat the case of the bibliography of the British National Library.

\section{Data and method}
\subsection{The British National bibliography}

National bibliographies are devised to include every publication in the country (Evans, 2005). The British Library is the legal depository of the United Kingdom and Ireland since 1662. The British National bibliography contains mentions of all books published since 1950.

In 2001, the British Library opened access to the British National bibliography (Deliot, 2014). It converted its documents from the library bibliographic format ''MARC21`` to the linked data format of the Resource Description Framework. This dataset has links to other open library datasets such as the Virtual International Authority File, Geonames or the Library of Congress Subject Headings. 

Each of the over three million British National bibliography records informs about the title, the author, the date and place of publication, and the subjects. 

\subsection{Network coincidence analysis}
The purpose of network coincidence analysis (Fisher 1924, 1928; Diaconis and Mosteler 1989; Escobar, 2015) is to detect which people, subjects, objects, attributes, or events appear simultaneously in different spaces, which are called  scenarios.  

$M$ events $X_j$, $j=1,\ldots,M$, are random variables recorded in each of the $N$ scenarios. $X_{j}=1$ if the $j$-th event occurs, $X_{j}=0$ otherwise. Two events are said to be ``coincident'' if they occur in the same scenario.

In the ``incidence'' matrix $X=(x_{ij})$, the rows  $i=1,\ldots,N$ stand for scenarios and the columns $j=1,\ldots,M$ for events. This matrix is binary, with elements $x_{ij}$ equal to $0$ or  $1$ indicating if the event $X_j$  occurs or not in  the $i$-th  scenario:

\begin{equation}\label{incidenceMat}
X = x_{ij} \quad {i=1,...,N; j=1,...,M.}
\end{equation}

The coincidence matrix $C=(c_{ij})_{i,j=1,...,M}$ is the symmetric $M\times M$ matrix
\begin{equation}\label{coincidenceFormula}
C:=X^{\top} X,\quad \quad \textrm{with} \quad c_{ij}:= \sum_{k=1}^N x_{ki}x_{kj}=c_{ji},
\end{equation}

\noindent{where $X^{\top} $ denotes the transposed matrix of $X$. Because only scenarios $k$ in which both events $X_i$ and $X_j$ occur ($x_{ki}=x_{kj}=1$) contribute to $c_{ij}$,  the  element $c_{ij}$  represents the total number of joint ocurrences of the events $X_i$ and $X_j$. The total number of scenarios in which $X_j$ occurs is $c_{jj}$.}

By definition, the two events $X_i$ and $X_j$ are independent of each other when the conditional probability $P(X_i | X_j)=P(X_i )$. Then the probability of recording both events  $X_i$ and $X_j$ is $P(X_i \cap X_j)=P(X_i )P(X_j)$. Two events  $X_i$  and $X_j$  coincide in probability if: 

\begin{equation}\label{coincidentProb}
c_{ij}>\frac{c_{ii}c_{jj}}{N},
\end{equation} 

\noindent{making $X_i$  and $X_j$ dependent of each other, or $P(X_i\cap X_j)>P(X_i)P(X_j).$}

We normalize the data using the statistical residuals $e_{ij}$ between the recorded and the expected values through the Pearson residual: 

\begin{equation}\label{equationPearson}
e_{ij}:=\frac{c_{ij}-\dfrac{c_{ii}c_{jj}}{N}}{({\dfrac{c_{ii}c_{jj}}{N}})^{\frac{1}{2}}},
\end{equation}

When non null, these residuals represent independent and coincident events. 

Haberman (1973) further divides $e_{ij}$ by the standard deviation of all residuals: 

\begin{equation} \everymath{\displaystyle}
d_{ij}:=\frac{e_{ij}}{((1-\frac{c_{ii}}{N})(1-\frac{c_{jj}}{N}))^{\frac{1}{2}}}, \quad i\neq j.
\end{equation}

The adjusted residuals, $d_{ij}$ are normally distributed with mean zero and standard deviation one. This allows us to test $d_{ij}=0$. With the entire population, it is no longer necessary to calculate probabilities. Thus we build the $M$ times $M$ adjacency matrix $A=(a_{ij})_{i,j=1,...,M}$ using the Haberman residuals $d_{ij}$ from all scenarios, following the rule:

\begin{equation}\label{adjacencyMat}
a_{ii}=0;\quad a_{ij}=\left\{\begin{array}{c} \mbox{$1$ if $d_{ij}>0$} \\ \\\mbox{$0$ if $d_{ij}\leq 0$} \end{array}\right. \quad i\neq j.
\end{equation}

With sample data, we compute the adjacency matrix with the probability that the adjusted residual $d_{ij}$ is non-negative.

Here the scenarios are the books mentioned in the British National bibliography. The events include subjects, authors, and publishers. In the incidence matrix $X$, the rows correspond to books and the columns to subjects, authors, or publishers. The coincidence matrix $C$ comprises the frequencies of those events in the diagonal and the frequencies of coincidences of two events elsewhere in the matrix. The adjacency matrix $A$ determines which events (authors, subjects, or publishers) coincide in the set of scenarios (books), with the Haberman residual $d_{ij}$ indicating the strength of that coincidence.

\subsection{Visual analytics}
Visualization is interactive: its purpose it to help make out patterns (Keim et al., 2008). Network graphs represent coincidences between events. A network graph $G=(V,E)$ represents a system made up of nodes $V=\lbrace v_{1}, v_{2},...,v_{m}\rbrace$ connected by edges $E=\lbrace e_{1}, e_{2},...,e_{l}\rbrace$ (Wasserman and Faust, 1994). We represent events as nodes and their coincidences as edges. The strengths of edges between connected nodes is given by Haberman residuals. The size of each node represents the frequency $c_{ii}$ of the event in the set of scenarios.

The spatial distribution of the nodes in a network depends on the method. Graph drawing algorithms are Fruchterman and Reingold´s (1991), Kamada-Kawai´s (1989), and multidimensional scaling (Kruskal, 1978).

\subsection{Software}
The R statistical software contains network coincidence analysis and the associated D3.js javascript visualization library in the netCoin R package (Escobar et al., 2017) available in the Comprehensive R Archive Network at \url{https://cran.r-project.org/package=netCoin}. 

NetCoin generates a Web page with network graphs of coincident events. Users load this Web page with a Web browser and interacts with the network through a control panel. They can customize the network for the location, color, shape, size of the nodes, and width and color of the edges. They can zoom and move the network.

\section{The case study}
After getting rid of irrelevant data, we retain frequent enough events. We calculate the strengths of edges between events, generate the interactive network graph, and remove edges having too low connection strengths.

\subsection{Topics of the British National bibliography over time}

Figure \ref{Timeline} shows that the total number of books catalogued per year in the British Library, has increased since 1950.

\begin{figure}[!ht]
  
  \centering
  \includegraphics[scale=0.5]{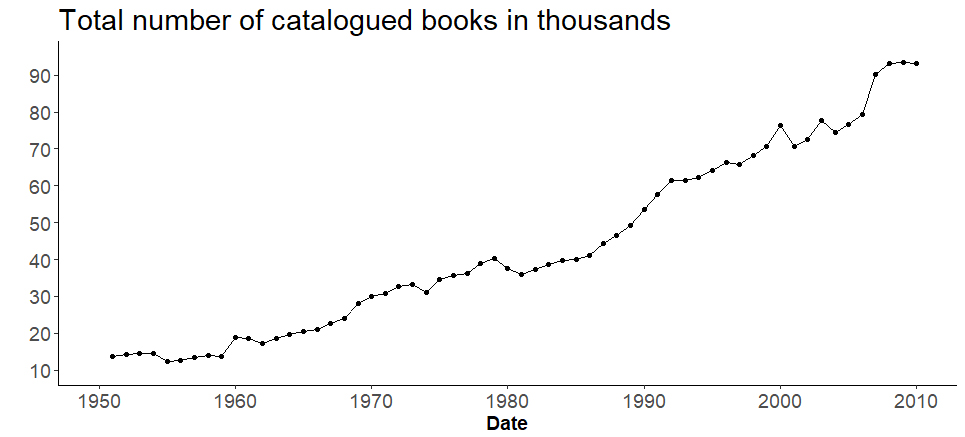} 
  \caption{Total number of books catalogued in the British National bibliography by year of publication.\label{Timeline}}
\end{figure}

We filter the scenarios by relevance. After discarding books with no date of publication or published before 1960, the dataset contains 2,816,615 books. Further discarding the books with no subject leads to 2,279,781 books. We classify all books published by decade of publication. The total set of scenarios comprises 287,233 subjects. We consider the most frequent 160 ones. We then produce the network graph in Figure \ref{ByDecade}, where decades are indicated with a cross and the size of the node represents the frequency of the event. The graph contrasts the 60s, 70s, and 80s on the left hand side of Figure \ref{ByDecade} to the 90s, 2000s, and 2010s on the right hand side. 

\begin{figure}[!ht]
  \centering
  \includegraphics[scale=0.8]{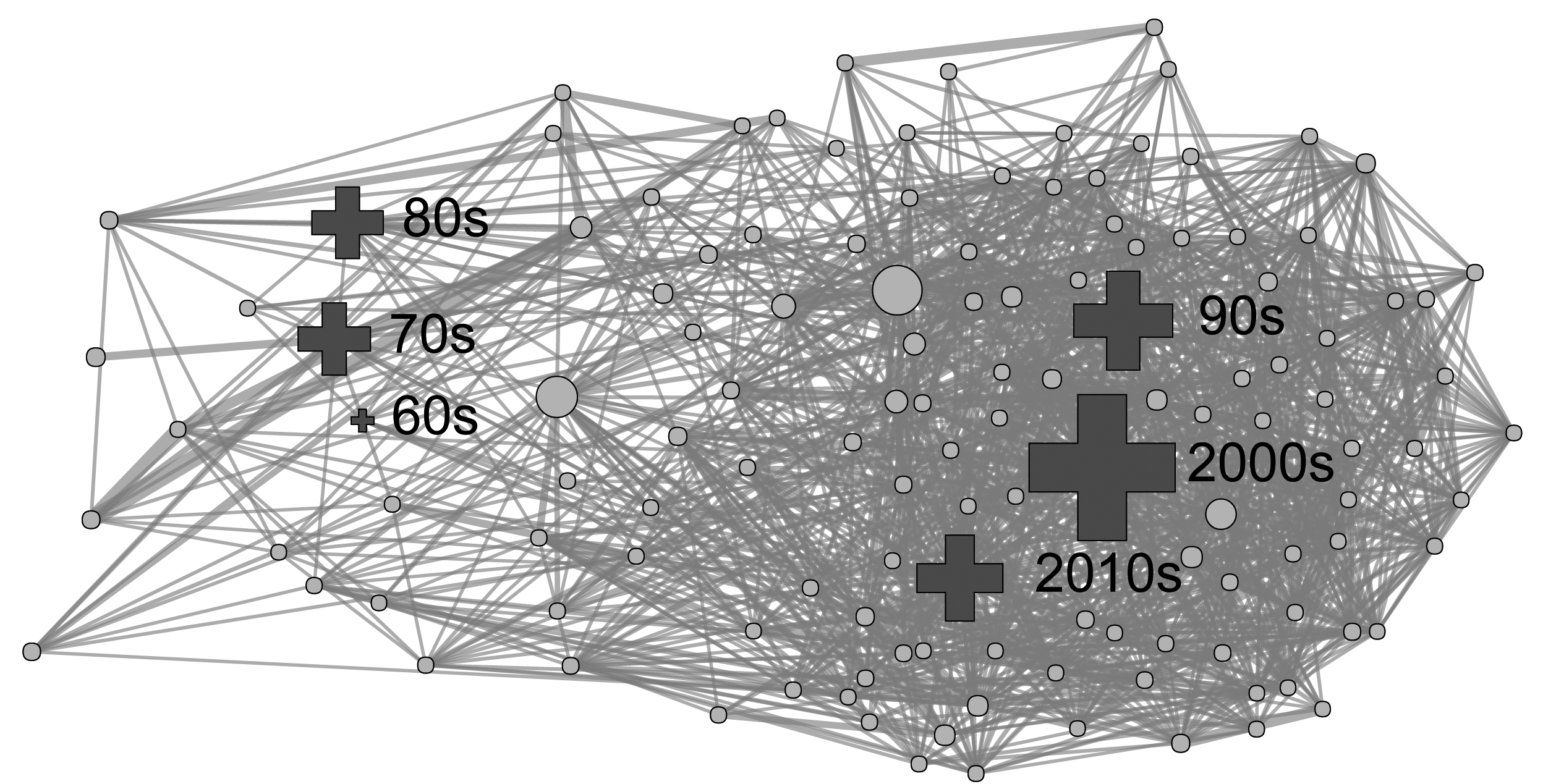} 
  \caption{Network of the main subjects in the British National bibliography by decade.\label{ByDecade}}
\end{figure}

\FloatBarrier
Figure \ref{ByDecade607080} presents the main subjects for the first three decades and Figure \ref{ByDecade900010} for the next three. The size of the nodes indicates the frequency of the events in the 2 million scenarios. These network graphs show that, although topics such as ``fiction in English'' keep a constant proportion across decades, others change. From the 60s to the 80s, the main subjects have ceased to be on Great Britain only, and, have gradually involved information technology, business, management, and social sciences, beside mathematics, physics, and geography.

\begin{figure}[!ht]
  \centering
  \includegraphics[scale=0.7]{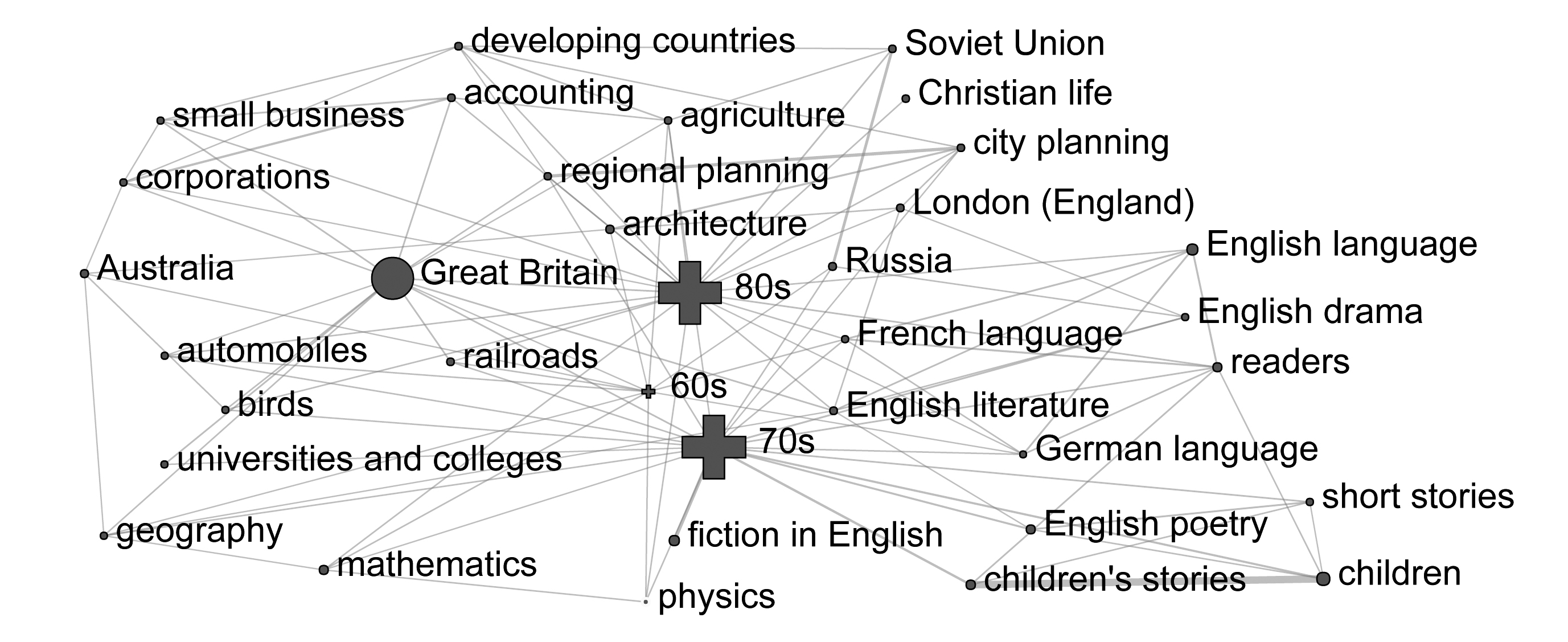} 
  \caption{Network of the subjects in the British National bibliography from 1960 to 1980.\label{ByDecade607080}}
\end{figure}

\vspace{5mm} 
\begin{figure}[!ht]
  \centering
  \includegraphics[scale=0.7]{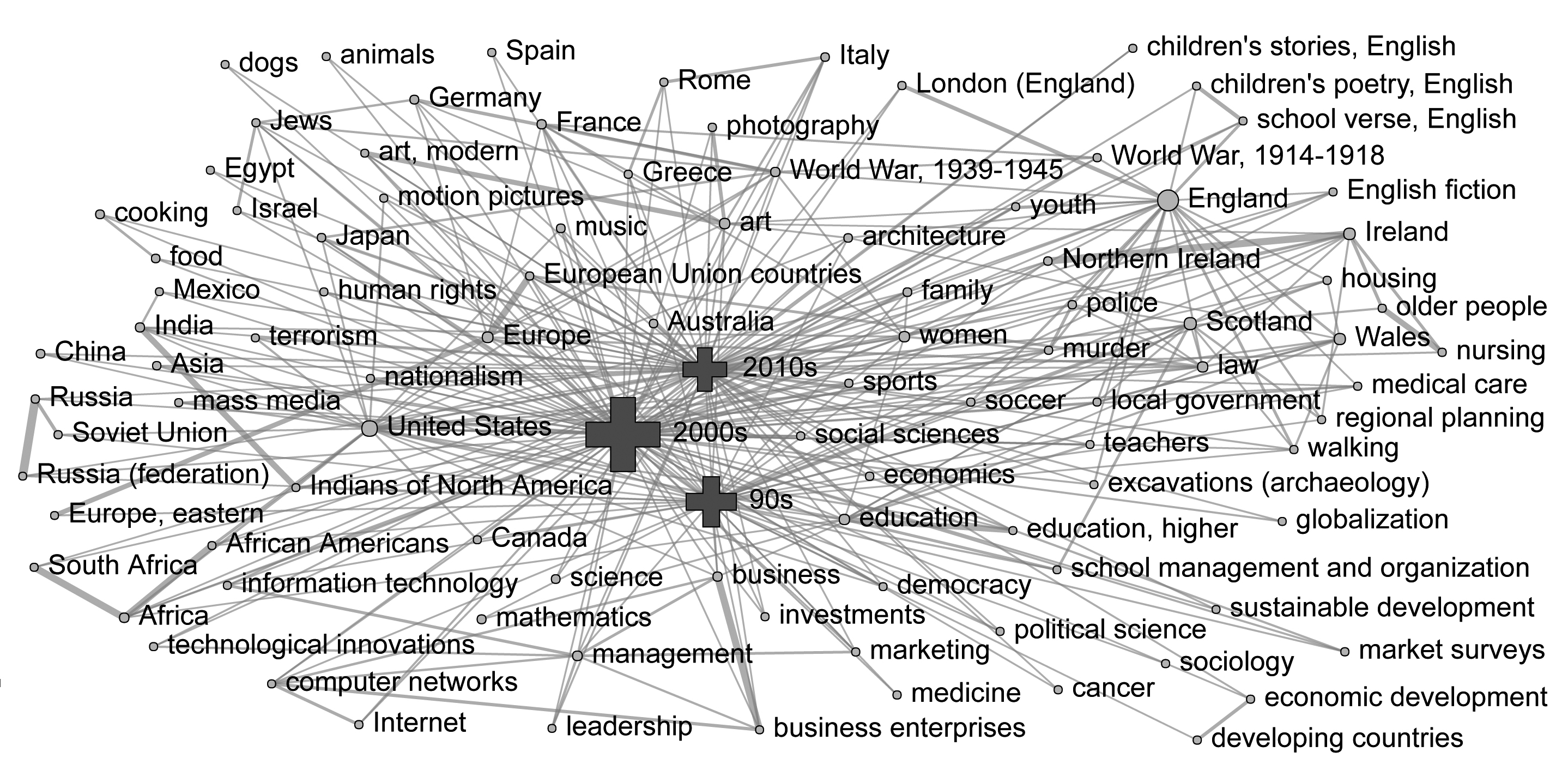} 
  \caption{Network of the main subjects in the British National bibliography from 1990 to 2016.\label{ByDecade900010}}
\end{figure}

Our data reduction has produced around 2 million scenarios and 300,000 events. The network graphs of figures \ref{ByDecade}, \ref{ByDecade607080}, and \ref{ByDecade900010} highlight the most prevalent and the strongest connections.  

\FloatBarrier
\subsection{Published themes}
The British National bibliography comprises 214,131 publishers. Figure \ref{FreqPubs} shows the publishers with the highest total number of books. How similar are their published themes\text{?}

\vspace{5mm} 

\begin{figure}[!ht]
  \centering
  \includegraphics[scale=0.7]{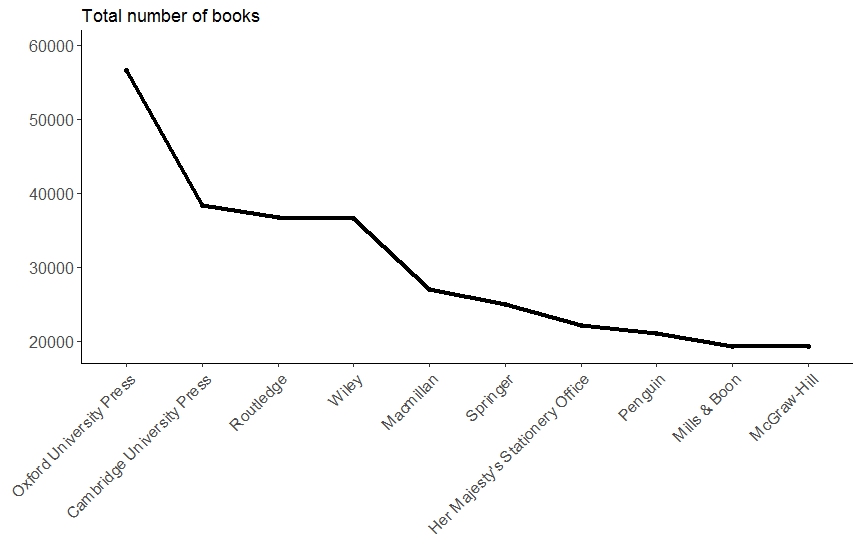} 
    \caption{Total number of books by publisher in the British National bibliography.\label{FreqPubs}}

\end{figure}

The data reduction starts with the selection of relevant scenarios, among the four most prolific publishers: Oxford University Press, Cambridge University Press, Routledge, and Wiley. This reduced set of scenarios comprises 168,294 books. 

We retain the four publishers and their 20\% most frequent topics.
For Oxford University Press, we retain 40 topics out of 20,126; for Cambridge University Press, 49 out of 17,972; for Wiley 52 out of 11,668; and for Routledge 80 out of 14,694. Because of common topics, the sum amounts to 157 topics, to which we add the four publishers considered as events to obtain 161 events. Figure \ref{ByPubFull} represents these topics together with the  publishers and their coincidences in a network graph using crosses for publisher nodes. Wiley is located away from the other publishers, which indicates that Wiley publishes relatively uncommon topics. The proximity of Oxford University Press, Routledge, and Cambridge University Press to one another indicates how much they share topics. Figures \ref{ByPubWiley} and \ref{ByPubOUPCamRou} highlight Wiley amidst its published topics, as an example.

\vspace{5mm} 

\begin{figure}[!ht]
  \centering
  \includegraphics[scale=0.8]{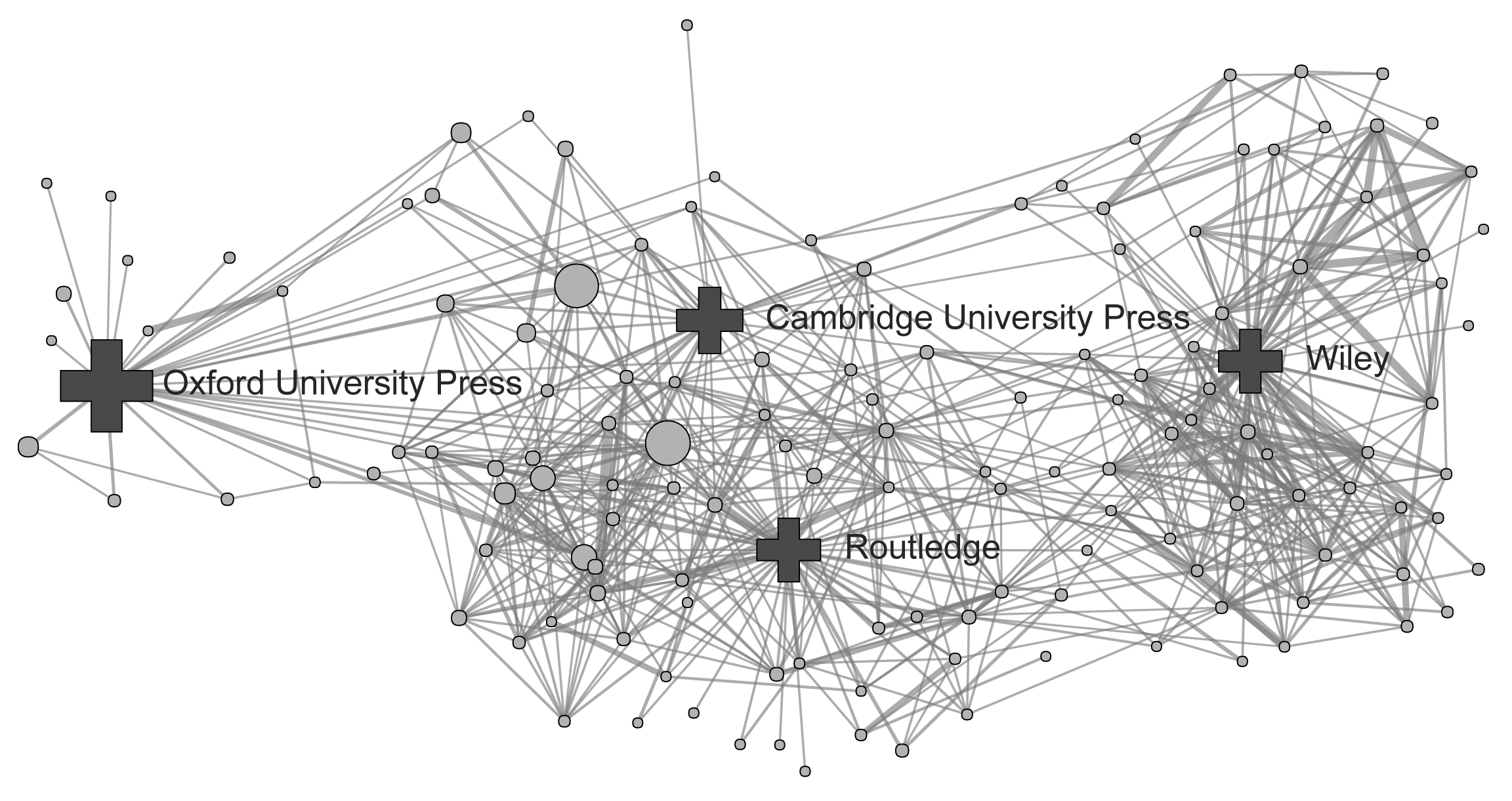} 
  \caption{Publishers and their topics.\label{ByPubFull}}
\end{figure}

\vspace{5mm} 

\begin{figure}[!ht]
  \centering
  \includegraphics[scale=0.9]{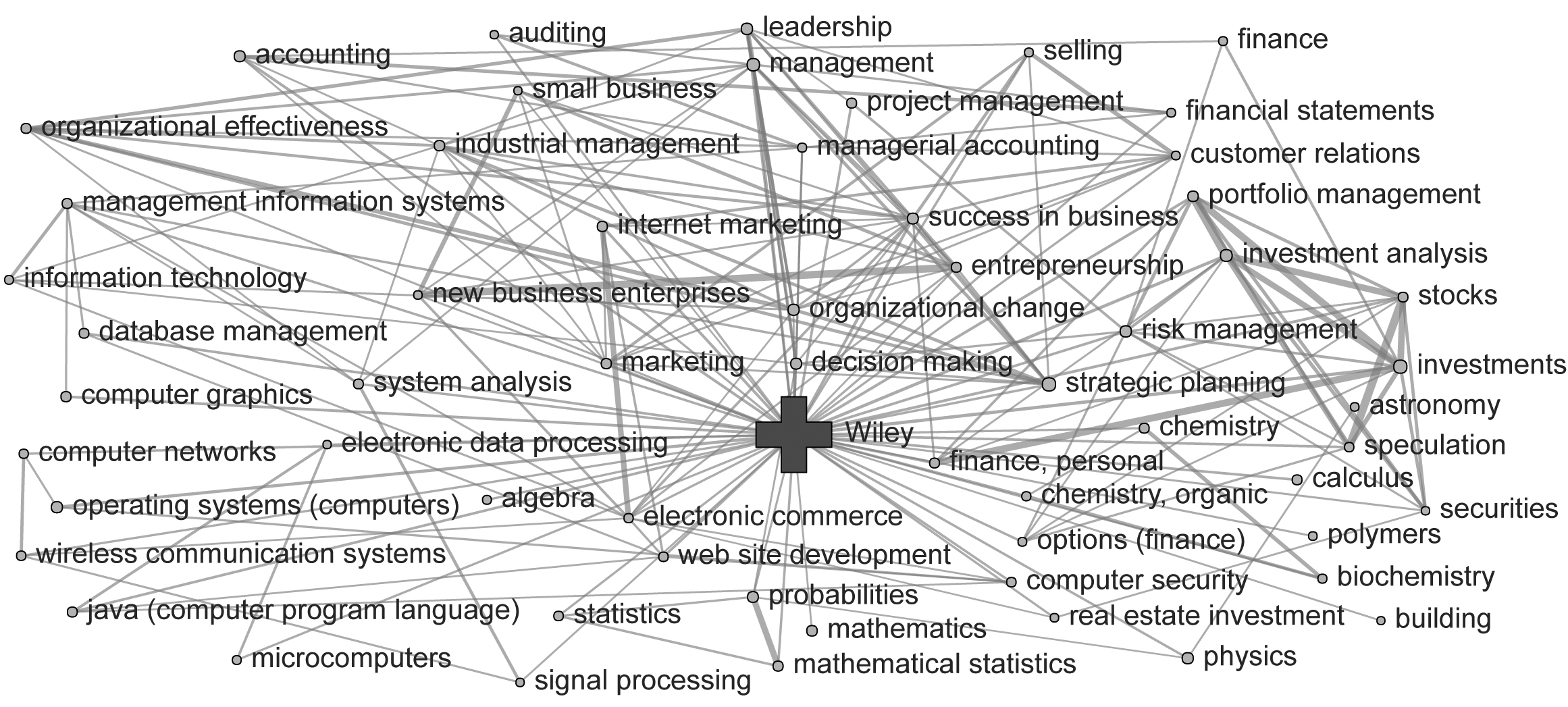} 
    \caption{Wiley and its topics.\label{ByPubWiley}}
\end{figure}

\vspace{5mm} 

\begin{figure}[!ht]
   \centering
  \includegraphics[scale=0.7]{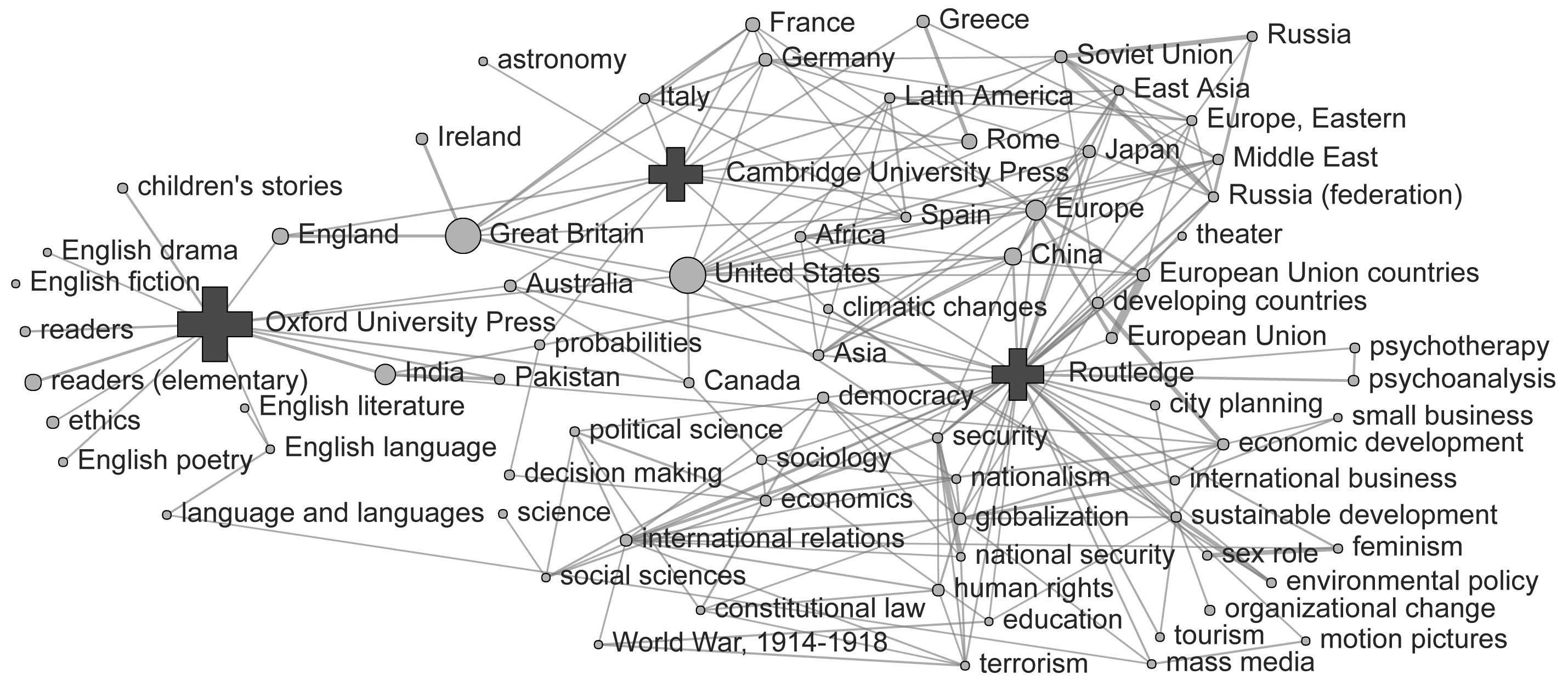} 
   \caption{Oxford University Press, Cambridge University Press, Routledge and their topics.\label{ByPubOUPCamRou}}
\end{figure}

\FloatBarrier
\subsection{The 100 best novels' authors of all times in the British National bibliography according to \textit{The Guardian}}

In October 2013, the newspaper \textit{The Guardian} published a list of what its editors considered to be the 100 best novels of all times (McCrum, 2013). How do their authors situate themselves in the British National bibliography?

The 100 authors are the ``events''. We reduce the data by selecting the books whose authors are mentioned in the 100 best novels by \textit{The Guardian}. They amount to 13,216 books from which we retain 6,613 books having clear topics, the ``scenarios''.
These books comprise 2,008 distinct subjects. We retain the 116 most frequent ones, amounting to more than half of all occurrences. For each author, we add images and geographical and topical information.

\begin{figure}[!ht]
  \captionsetup{width=0.8\textwidth}
  
  \includegraphics[scale=0.65]{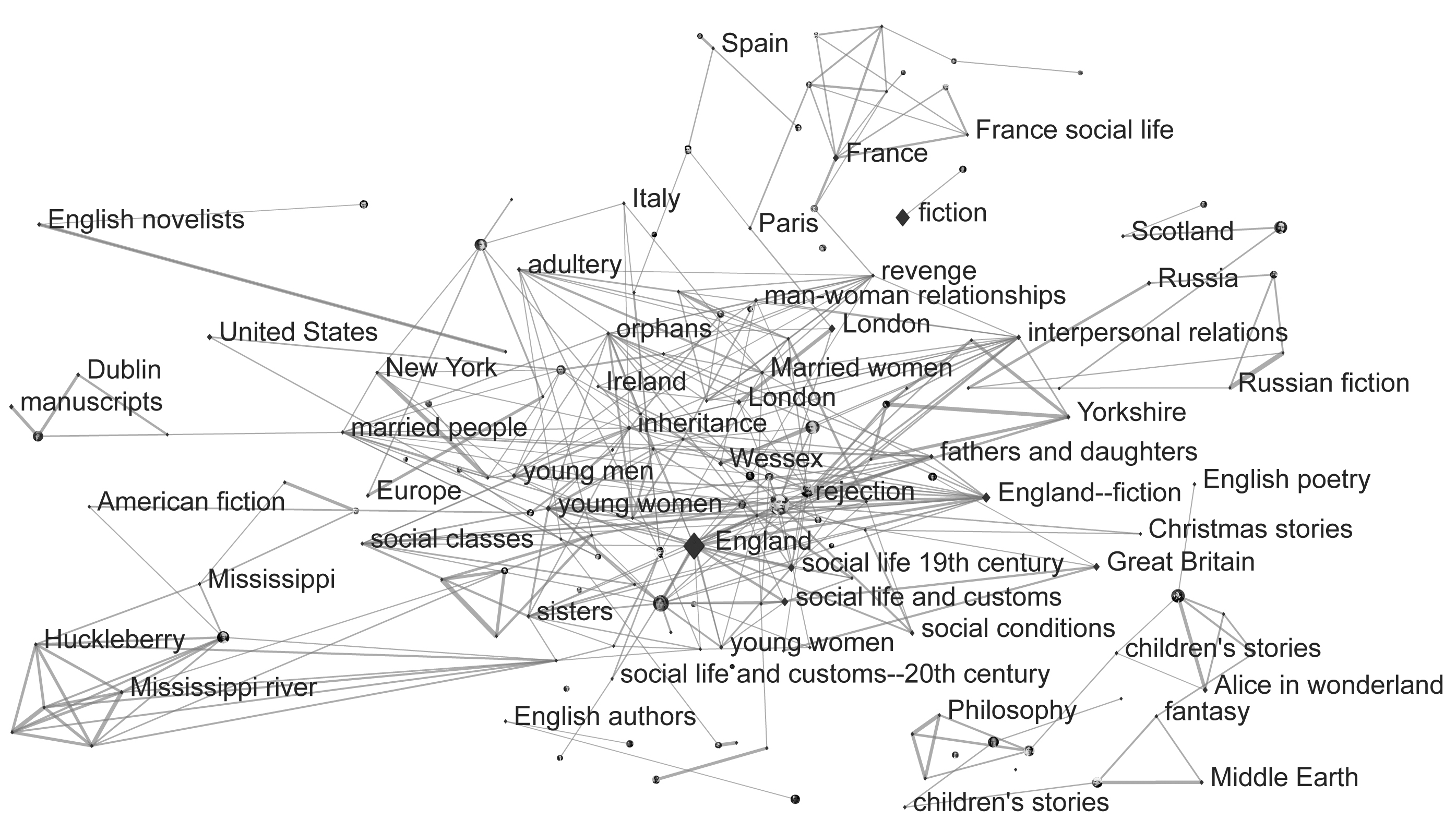} 
  \centering
  \caption{The 100 best novels' authors according to \textit{The Guardian} in the British National bibliography witht their main topics.\label{GuardianFull}}
\end{figure}

The resulting network in Figure \ref{GuardianFull} highlights an English centric view of literature, with ``England'' as the most frequent event.

\vspace{5mm} 

\begin{figure}[!ht]
  \captionsetup{width=0.8\textwidth}
  \includegraphics[scale=0.6]{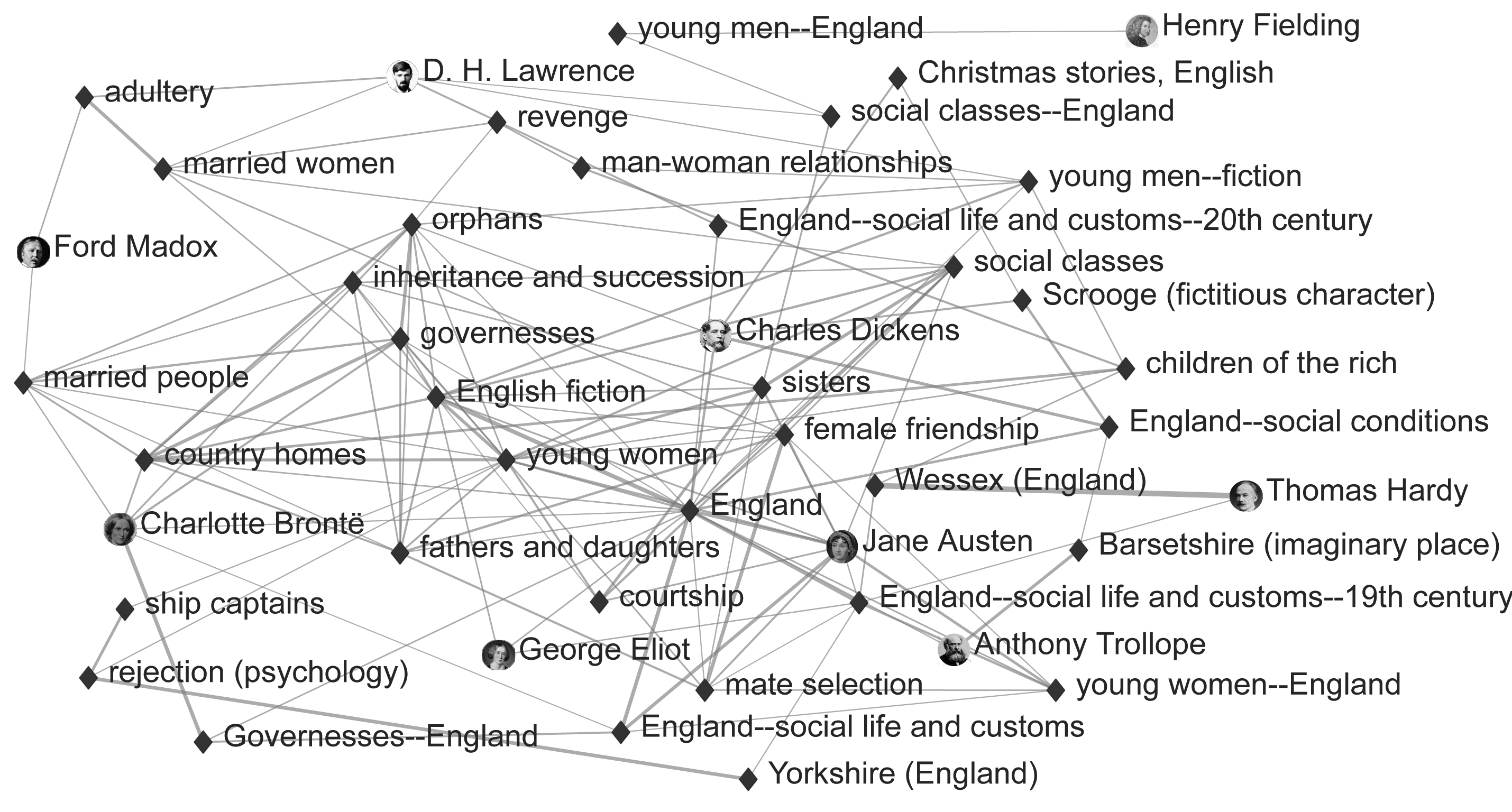} 
  \centering
  \caption{The 100 best novels' authors in the British National bibliography surrounding the ``England'' node.\label{GuardianEngland}}

\end{figure}

Authors around the node of ``England'' include mostly romantic and realistic British authors such as Charles Dickens, Jane Austen, and Anthony Trollope, and treat fiction, family and friendships, and social life in England. 

\FloatBarrier
\section{Conclusion}
We showed that collections prepared by libraries can contribute to Big Data. Network coincidence analysis combines statistical methods and social network analysis to reduce the total number of events, measure relationships, and delineate trends. 
Our treatment of a bibliography has shown how to extract insights from more than 3 million records, 800,000 person names, and 300,000 subject headings. 

\section*{Acknowledgement}
This article was financed by the Spanish Ministry of Economy and Competitiveness and the European Regional Development Fund (FEDER) (research project number CSO2013-49278-EXP).

\end{document}